\documentclass[12pt,preprint]{aastex}
\usepackage{subfigure}
\usepackage{graphicx}

\newcommand\etal{{\it et al.\/}}

\newcommand{\lephare}{{\it Le~Phare\/} }
\newcommand{\lephares}{{\it Le~Phare}'s\/ }
\newcommand{\Lephare}{{\it Le~Phare}\/}
\newcommand{\sub}[1]{\mbox{\footnotesize{#1}}}

\begin{document}

\slugcomment{CVS \$Revision: 1.9 $ $ \$Date $ $}

\title{Photometric Redshift Biases from Galaxy Evolution}

\author{C. Jonathan MacDonald \& Gary Bernstein}
\email{cmac@physics.upenn.edu, garyb@physics.upenn.edu}
\affil{Department of Physics \& Astronomy, University of Pennsylvania, 
209 S.\ 33rd St., Philadelphia, PA 19104}

\begin{abstract}
Proposed cosmological surveys will make use of photometric redshifts of galaxies that are significantly fainter than any complete spectroscopic redshift surveys that exist to train the photo-z methods.  We investigate the photo-z biases that result from known differences between the faint and bright populations: a rise in AGN activity toward higher redshift, and a metallicity difference between intrinsically luminous and faint early-type galaxies.  We find that even very small mismatches between the mean photometric target and the training set can induce photo-z biases large enough to corrupt derived cosmological parameters significantly.  A metallicity shift of $\sim0.003$~dex in an old population, or contamination of any galaxy spectrum with $\sim 0.2\%$ AGN flux, is sufficient to induce a $10^{-3}$ bias in photo-z.  These results highlight the danger in extrapolating the behavior of bright galaxies to a fainter population, and the desirability of a spectroscopic training set that spans all of the characteristics of the photo-z targets, {\it i.e.} extending to the 25th mag or fainter galaxies that will be used in future surveys.

\end{abstract}


\section{Introduction}
Photometric redshifts have been used extensively in studies of galaxy
evolution, particularly for sources that are too faint for feasible
spectroscopic detection of absorption or emission lines.  Photo-z's
are just beginning to find application to measurements of cosmic
structure, particularly in the determination of source redshifts for
weak gravitational lensing surveys \citep[e.g.][]{Combo17,Cosmos}.  
Projects in development plan to
measure photometric redshifts for $\sim10^9$ galaxies, far more than
can be practically determined spectroscopically. 

Yet the cosmological
measures that weak-lensing surveys have as goals are extremely
sensitive to biases or other miscalibrations of the photo-z scale:
biases as small as $\approx10^{-3}(1+z)$ will overwhelm the
statistical errors on cosmology in these experiments \citep{HTBJ}.
The photo-z technique has never been subject to such demands on its
accuracy, so we need to examine carefully the strategies for achieving
them.

A photo-z algorithm must somehow be taught to convert broadband
fluxes into redshifts.  This can be done by fitting to theoretical or
empirical templates for galaxy spectra \citep[e.g.][]{Benitez}; or by completely empirical
machine-learning techniques, {\it e.g.} neural nets \citep{Annz}, that are trained
with a sample of galaxies of known spectroscopic redshift.  There is a
difficulty, however, in that many imaging surveys intend to measure
photo-z's from galaxies that are too faint to obtain the spectroscopic
information needed to form templates or obtain spectroscopic training
redshifts.  One must therefore take on some degree of faith the
assertion that a photo-z algorithm trained on brighter galaxies will
return sufficiently unbiased redshifts for galaxies beyond the
spectroscopic limit. 

We investigate in this paper the sizes of biases that we might expect
to arise because of differences between the faint galaxy population
and the brighter spectroscopic population.  In particular:
\begin{itemize}
\item Fainter galaxies will tend to be at higher redshifts, where
  active galactic nuclei (AGN) are more frequent or brighter than at
  low redshift.  Furthermore, optically-based spectroscopy is
  inefficient in the ``desert'' between $1.5\lesssim z \lesssim 3$,
  where quasar activity peaks.  We can ask, therefore, whether the
  different AGN characteristics of the fainter galaxies could bias the
  photo-z's trained on low-z galaxies.
\item For elliptical galaxies, there is a well-known color-magnitude
  relation. As we move to lower luminosity galaxies in this red sequence, stellar populations traverse a path in the 
age-metallicity plane.  Stellar populations of
the luminous galaxies of different Hubble types that would typically be used for photo-z training will not cover the same locus in the age-metallicity plane.
Would photo-z's trained on bright galaxies produce biased
  redshifts for fainter ellipticals?  We investigate this by examining the metallicity dependence of photo-z results for old stellar populations when training on the bright Hubble sequence. There are additional complications from the age-metallicity degeneracy. Accordingly, we also estimate the age dependent behavior of photo-z biases.
\end{itemize}

In the Rumsfeld parlance, these are ``known unknowns'' of the faint
galaxy population: a photo-z algorithm might be designed which could
compensate for these two differences between faint and bright
galaxies.  But we will use a photo-z algorithm that is ignorant of
these effects, so that they can serve as models for ``unknown
unknowns'' which might bias faint photo-z estimates.  In this way we
might learn how safe or dangerous it is to extrapolate training sets
or templates to fainter magnitudes.

\section{Photo-z fitting technique and template sets}
Probing AGN contamination biases necessitates the preparation of an appropriate galaxy spectrum and the use of fitting software to determine the photometric redshift $z_p$. Contaminated galaxies are modeled by introducing a QSO spectrum as an impurity to the spectral energy distributions (SEDs) of galaxies. The {\Lephare}\footnote{{\url{http://www.oamp.fr/people/arnouts/LE\_PHARE.html}}} software package is used to compute the photometric redshifts. \lephare compares theoretical and known source magnitudes, estimating $z_p$ via $\chi^2$ minimization \citep[{\it cf}.][]{steve1,steve2}.

\lephare fits to a magnitude catalog of template galaxies. In this case, the catalog consists of 4 known galaxies from \citet{CWW}[CWW] and 3 simulated galaxies from \citet{BC96}[BC96] as well as linear interpolations (via flux) of these. The CWW galaxies are early types, Sbc and Scd spirals, and irregulars, while the BC96 galaxies are $2$, $0.5$, and $.05$\,Gyr in age. Altogether, including interpolations, the catalog is composed of 72 models. We investigate two sets of filters. The first includes 4 Hubble Deep Field North (HDFN) filters (F300W, F450W, F606W, and F814W) and three near-infrared filters (Jbb, H, and K) included in the \lephare software package. The second set is the 5 Sloan Digital Sky Survey (SDSS) filter set (u, g, r, i, and z). \lephare compiles this catalog by calculating the magnitudes for each galaxy SED with each filter at redshifts from $0$ to $6$ in increments of $.04$.  Note that the template galaxies contain no AGN signal.

A second input catalog describes the galaxies for which $z_p$ is to be estimated. We synthesize AGN-contaminated magnitudes by: making a QSO template spectrum; normalizing this template to the galaxy SEDs; contaminating the \lephare templates with a specified fraction of AGN flux; then calculating apparent magnitudes at various redshifts and AGN contaminations.

A template QSO spectrum over a wide wavelength span is needed in order to synthesize AGN contaminated SEDs. \citet{VB} and \citet{GHW} provide suitable QSO template spectra; the former covers a broad UV-NIR (800.5\AA - 8554.5\AA) wavelength range while the latter extends further into the NIR (5801\AA - 35095.4\AA). We scale the Glikman spectrum to match the Vanden Berk spectrum at 847~nm, where both are well measured, and then concatenate the two.

The QSO contamination is specified by the fraction $0\le h\le 0.2$ of the flux contributed by the QSO.  Specifically, for a chosen galaxy template $g(\lambda)$ and our template QSO spectrum $q(\lambda)$ we first determine the normalization $N$ which yields
\begin{equation}
\int^{\lambda_2}_{\lambda_2} g(\lambda)\,d\lambda = N\int^{\lambda_2}_{\lambda_1} q(\lambda)\,d\lambda,
\end{equation}
where the integration bounds $\lambda_1 =80$~nm to $\lambda_2=3.5\,\mu$m span the range over which both spectra are known.
For a desired contamination $h$, the galaxy rest-frame SED is taken as
\begin{equation}
f(\lambda) = (1 - h)\,g(\lambda) + N\,h\,q(\lambda).
\end{equation}
Apparent magnitudes in the chosen filter sets are calculated using the standard formulae for redshifted sources.

We use the same QSO template spectrum for all levels of contamination.  In reality, low-luminosity AGN will have different spectra than bright QSOs, {\it e.g.} weaker broad lines, however the effect of contamination on photo-z's that we derive will still be crudely correct.

The catalog of AGN-contaminated galaxies contains
7 galaxy types---the 4 CWW templates (of type E, Sbc, Scd, and Irr) and the 3 BC96 simulated star forming galaxies (denoted I2, I05, and I005 for their ages)---at redshifts in the range $0 \leq z \leq 3$ in steps of $.1$ for SDSS filters, and in steps of $.1$ from $0 \leq z \leq 1$ and steps of $.2$ for $1 \leq z \leq 3$ for the HDFN filters, with AGN contaminations in the range $0 \leq h \leq 0.2$ in $0.01$ increments.

We run \lephare to obtain estimates for $z_p$. The quantity of interest is the redshift bias between the nominal and photometric redshifts, to wit, $\Delta z = z_p - z$.

\section{Bias from AGN contamination}
\begin{figure}[t!]
	\epsscale{.323}\plotone{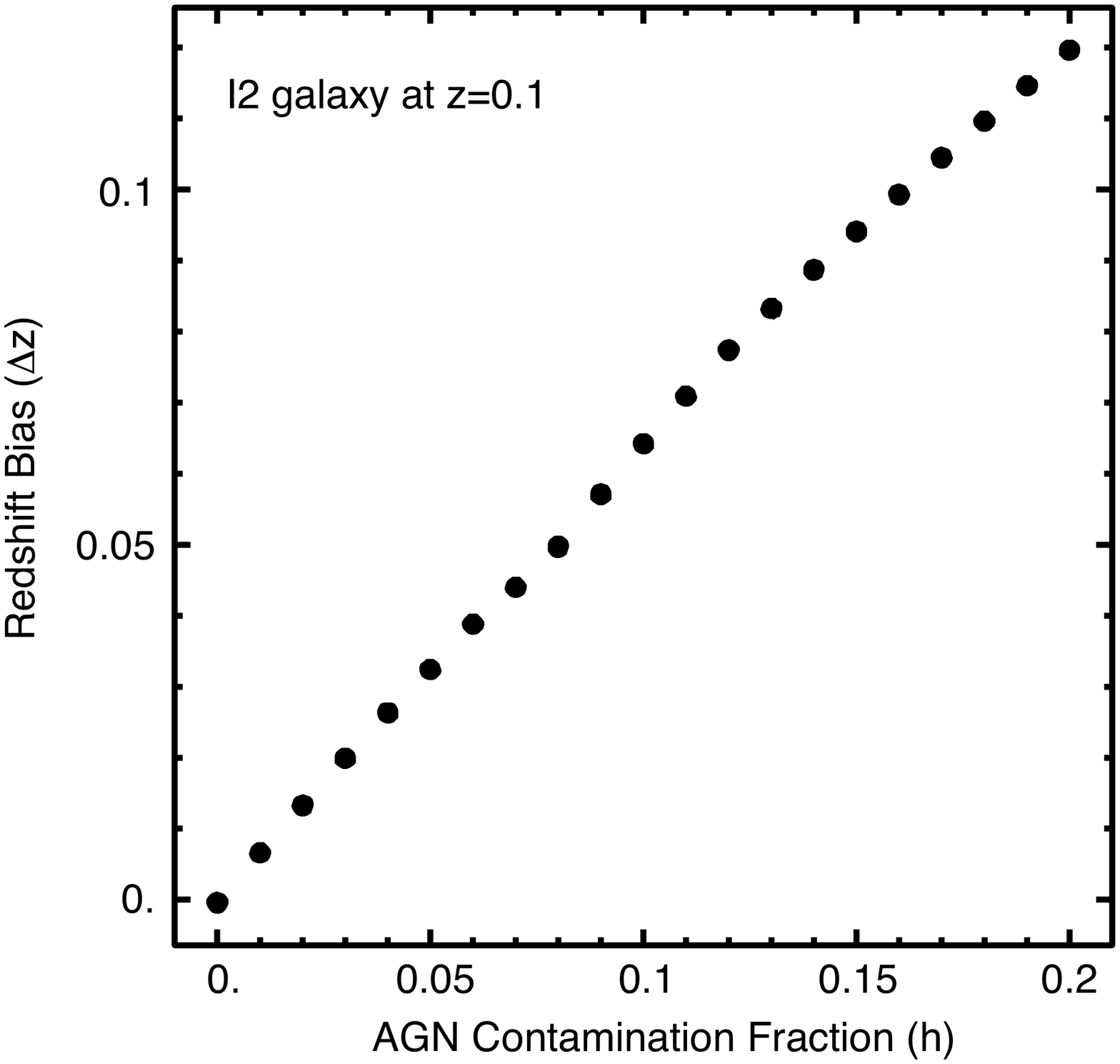}
	\epsscale{.323}\plotone{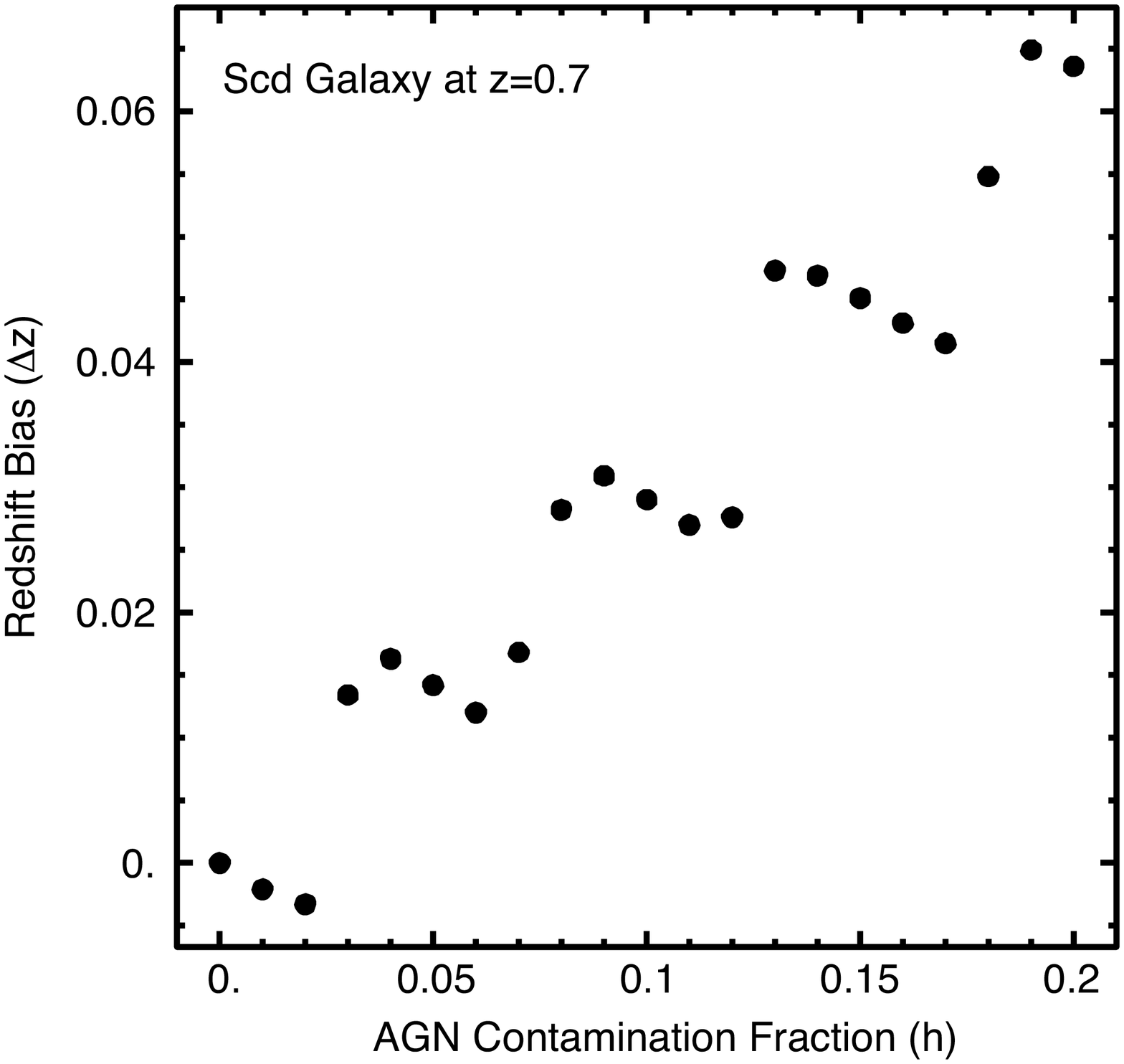}
	\epsscale{.323}\plotone{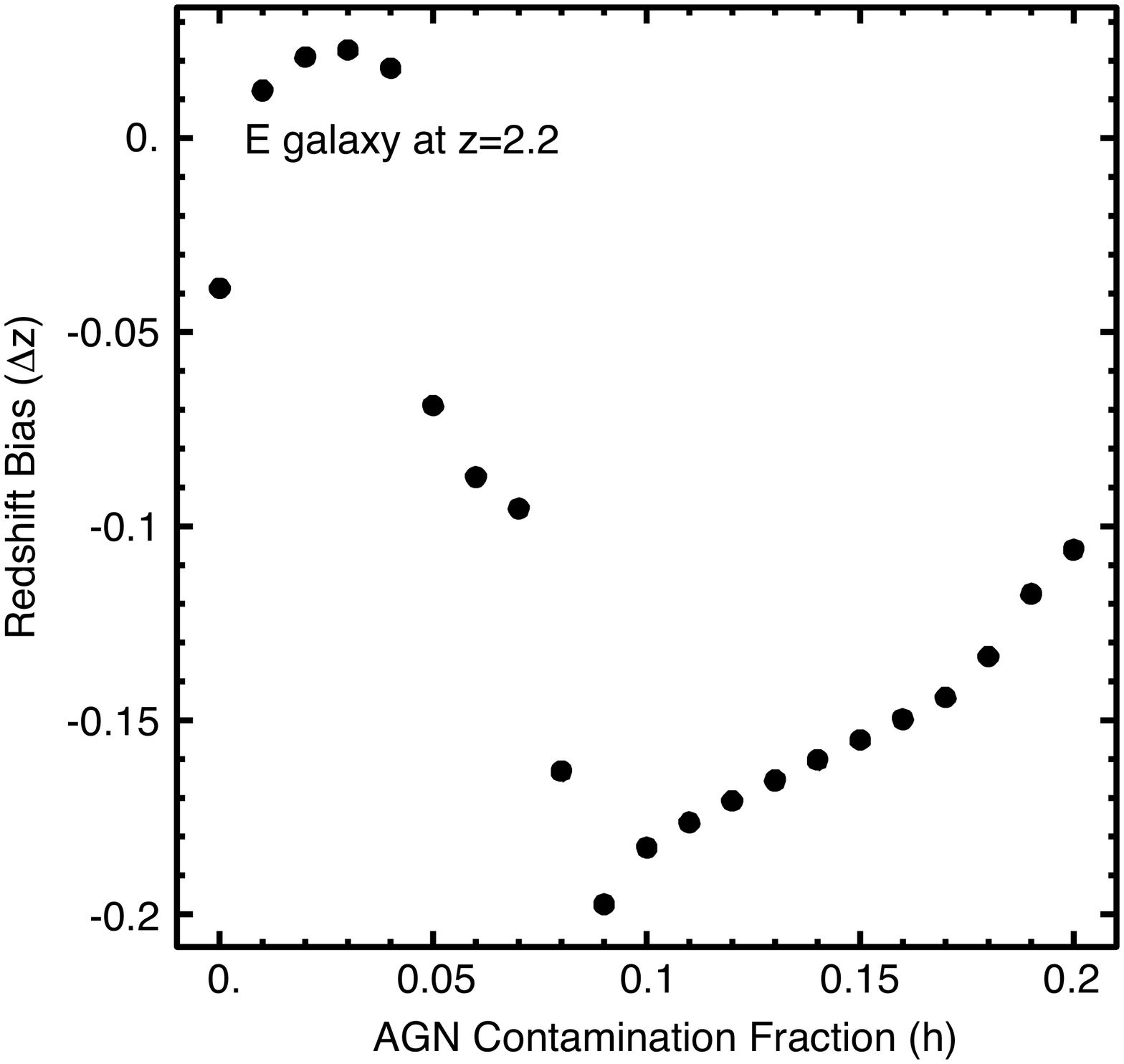}
	\caption{\small Plots of $\Delta z$ vs. $h$ with various behaviors.}\label{gbu}
\end{figure}

All galaxy types using either filter set have a non-trivial bias $\Delta z$ induced by AGN contamination.  We first examine how $\Delta z$ depends on contamination fraction $h$, at fixed $z$ and galaxy type. Figure~\ref{gbu} illustrates the range of behaviors found. The I2 galaxy shows a simple linear bias-contamination response with a well-defined $d(\Delta z)/dh$. In cases  like the Scd galaxy at $z=0.7$, the linear trend gains steps or wiggles, which we believe is induced by the discrete steps in galaxy type that \lephare takes when finding a best-fit template. In this case we define $d(\Delta z)/dh$ by a least-squares fit through all the data points.
Finally there are cases like the high redshift E-type in which the best-fit photo-z becomes ``catastrophically'' incorrect above some contamination level.  Here the AGN light has moved the galaxy toward some new part of the color manifold of the templates rather than inducing a small differential change.  We define 
$d(\Delta z)/dh$ by manually selecting a region of $h$ that is below the catastrophe and fitting a slope to this region.  We essentially ignore the catastrophic failures because we are more interested in biases induced by gradual evolution in the AGN fraction.  We will focus henceforth on the slope 
$d(\Delta z)/dh$ in the vicinity of $h=0$.


\begin{figure}[b!]
	\epsscale{.406}\plotone{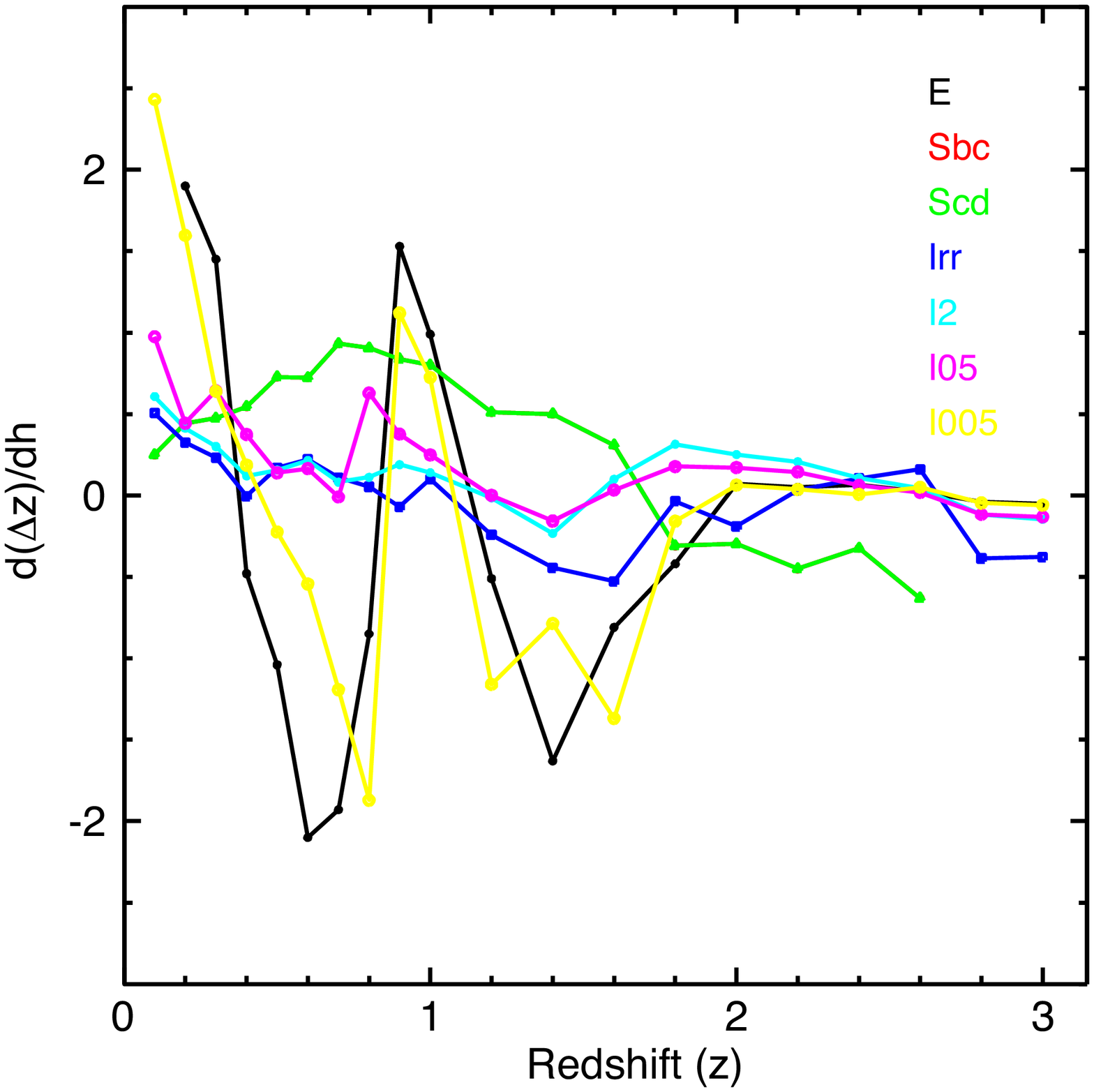}
	\epsscale{.406}\plotone{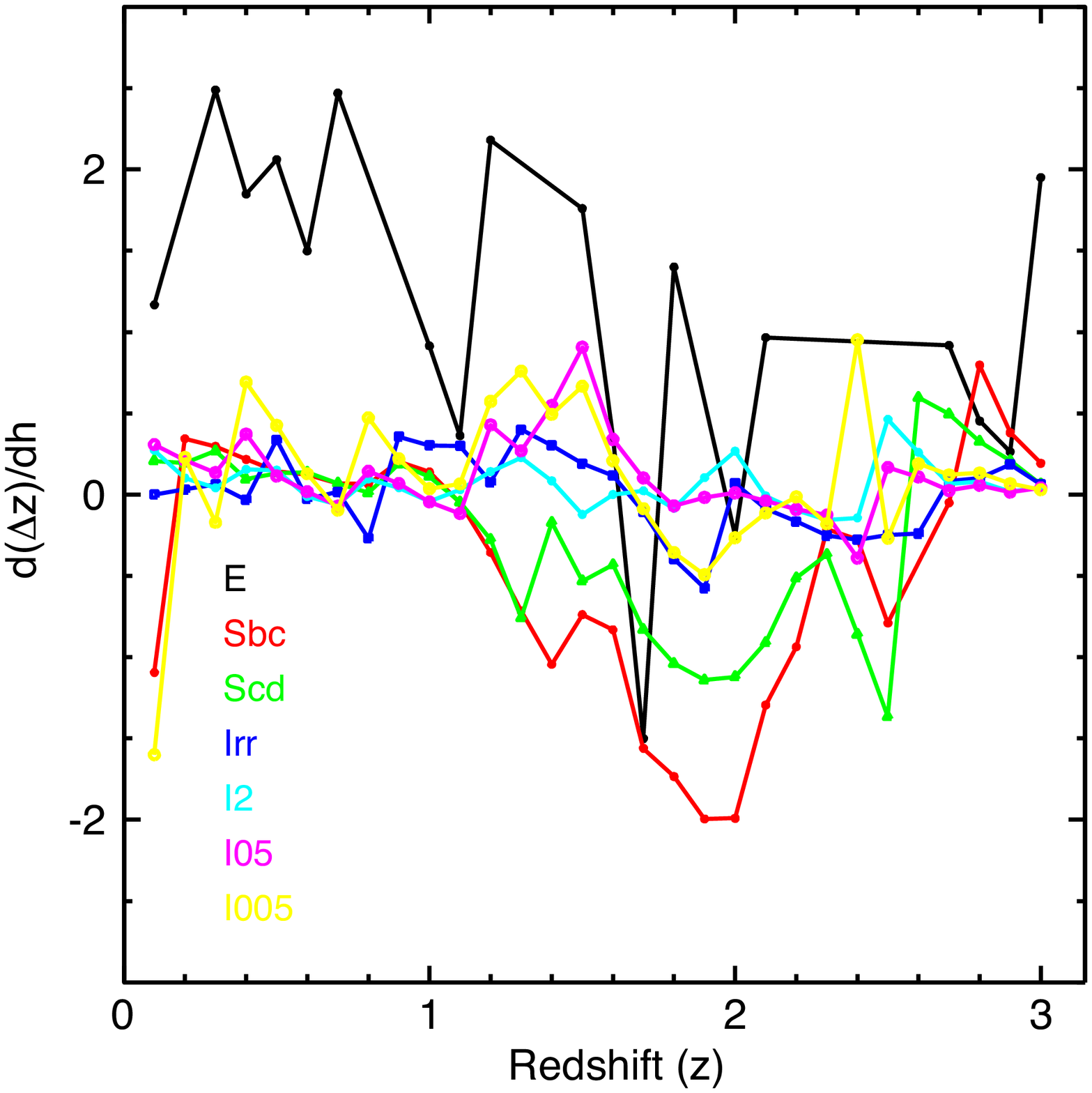}
	\caption{\small The response $d(\Delta z)/dh$ of photo-z bias to AGN contamination is shown for 	two filter sets (HDFN on the left and SDSS on the right). The bias response is plotted vs 				redshift for different galaxy types. For clarity, in the plot we require $|d(\Delta z)/dh| < 			2.5$ which cuts out some catastrophic errors.}\label{dzdp}
\end{figure}
Figure~\ref{dzdp} shows $d(\Delta z)/dh$ from the low-contamination domain in both filter sets. The different templates show a variety of behaviors, and some are erratic due to discreteness in the \lephare fitting, particularly with the SDSS filter set, which offers poor photo-z constraints in some domains due to lack of NIR coverage.

The important point, however, is that the bias responses $d(\Delta z)/dh$ are $\approx\pm0.5$ over most of the color and galaxy-type domain.  {\em A 1\% contamination with AGN light causes a photo-z bias of $\Delta z\sim0.005$} if the photo-z algorithm is using uncontaminated templates.  Application of photo-z's to cosmic-shear measurements require that biases be held below $\sim0.001$, which means that the mean AGN contamination of the target sample would need to be determined to $\ll1\%$ in order for the resultant photo-z biases to make insignificant impact on the inferred cosmology.  We will not know the faint-galaxy AGN fraction to anywhere near this precision, so this would be a major problem.

Of course we are using a naive photo-z estimator: one could include a QSO template in the fitting process (as in the option \lephare offers, or as described in \cite{WLF}, which includes object classification), or attempt to construct an empirical training set of bright galaxies that spans the range of AGN contamination of the faint set. This would greatly reduce the photo-z bias in our simple simulated catalog. The behavior of real AGN is, however, much more complex than our simple single QSO template, and it is likely that the faint population has different AGN spectra than the bright galaxies that might serve as templates. We simply point out that photo-z biases can be very sensitive to AGN light, and it is risky to extrapolate AGN characteristics to a new population.

\section{Age-Metallicity Models}
We use stellar population models to estimate the redshift bias induced by a metallicity mismatch between target galaxies and \lephares templates (\lephare uses the same extended CWW catalog used above). The galaxy evolution simulation code GALAXEV\footnote{\url{http://www2.iap.fr/users/charlot/bc2003/}} is used here to generate a magnitude catalog for galaxies with near-solar metallicity with a $z=0$ ages of 10\,Gyr. We use the standard model as described in \citet{BC03}[BC03]. Specifically, the galaxy models employed here use the STELIB and BaSeL spectral libraries, a simple stellar population (SSP) star formation rate, the Chabrier initial mass function, and the 1994 Padova evolutionary tracks (per recommendation of the code's authors).

GALAXEV produces a table of magnitudes and ages at various redshifts for a given galaxy model and filter set. The same SDSS and infrared supplemented HDFN filters from the AGN contamination study are used to calculate these magnitudes. GALAXEV produces colors for metallicities $Z/Z_{\odot}=0.2, 0.4, 1.0,$ and 2.5 (models m42, m52, m62, and m72 in GALAXEV). We also produce colors for metallicities 0.3, 0.7, and $1.75Z_{\odot}$ by interpolating between these four GALAXEV spectra.

We further employ GALAXEV to investigate biases due to age misestimates. We generate solar metallicity models with z = 0 ages of 11, 9, and 8 Gyr. Note that the fiducial model here is still the 10\,Gyr old solar metallicity galaxy, but in contrast to what is done in the metallicity study, we now hold the metallicity fixed and vary the age.

\section{Bias From Metallicity Indeterminacy} 
Calculating the bias due to a small metallicity mismatch requires removing the fiducial bias present resulting from inherent differences between the GALAXEV spectra and \lephares catalog of templates. At each redshift, we take an elliptical galaxy with solar metallicity as our benchmark. The solar bias, $\delta z_{\odot}$, is defined as the difference between the solar model's redshift and \lephares estimate:~$\delta z_{\odot}=z_{\odot,\mbox{\tiny mod}}-z_{\odot,\mbox{\tiny phot}}$. For any particular metallicity, the bias due to metallicity mismatch is then $\Delta z = z_{\rm phot} - z_{\rm model} - \delta z_\odot$.  Hence by definition we assign zero metallicity bias to a solar-metallicity population, for any chosen filter set, population age, and redshift.

\begin{figure}[!hb]
	\epsscale{.406}\plotone{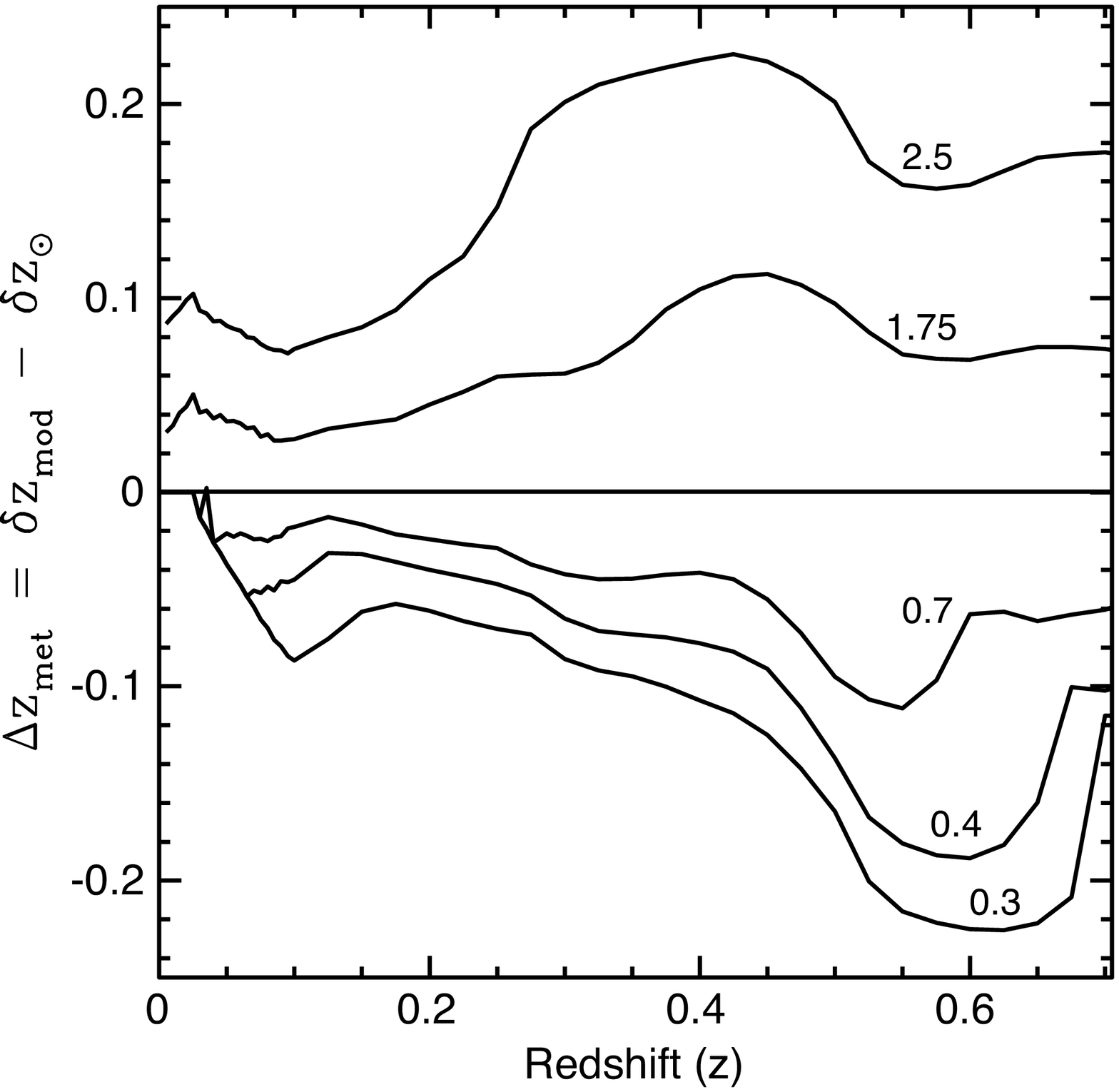}
	\epsscale{.406}\plotone{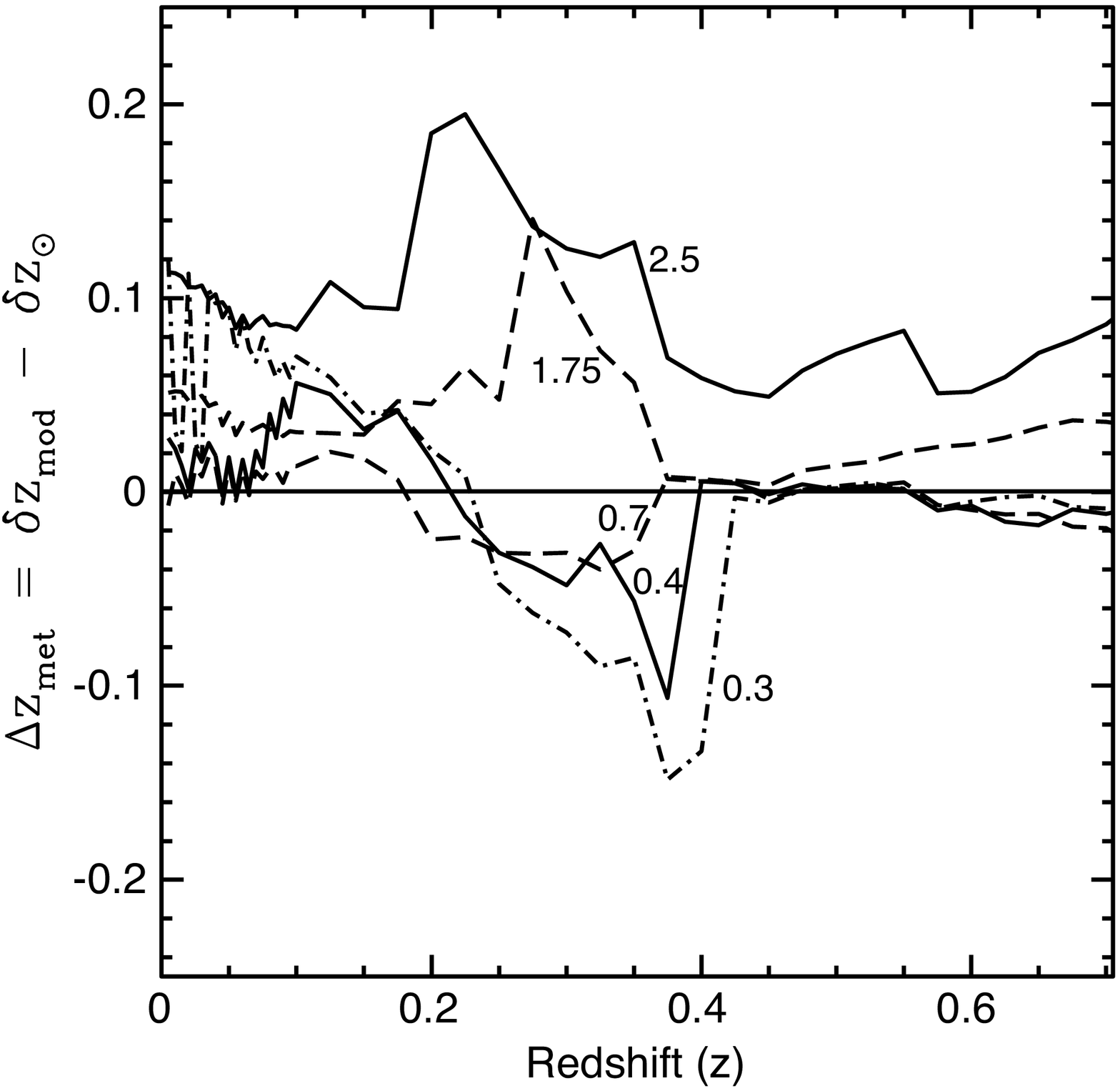}
	\caption{\small Metallicity-induced biases are shown from rest-frame to z=.7 for HDFN (left) and   SDSS (right)	filters. Each model's metallicity is shown on the plot in solar units. The dashed 		and dotted lines serve distinguish between the models in the crowded regions. The smoother    		behavior of the HDFN+IR plot is attributable to the supplementary IR bands---the HDFN data acquire jagged profiles when the analysis is done sans the IR bands. Indeed, these plots (not 		shown) resemble the SDSS plots; while the overall range of the bias decreases, the curves 				behave more erratically.}\label{mzb}
\end{figure}

Figure~\ref{mzb} illustrates the biases from metallicity offsets as a function of redshift between $z=0$ and $z=.7$. Five models are present in the plot, each is 10\,Gyr old at $z=0$ with metallicities 0.2--2.5$Z_\odot$. As the models get more metal rich, the redshift bias increases, that is, the photometric redshift gets larger than the model's redshift. In the redshift range that was studied, the super- and sub-solar models both had biases of up to $\pm .2$ for metallicities $\approx\pm0.5$~dex away from solar. This suggests that biases of $\approx0.001(1+z)$ would result from shifts of only $\sim0.003$~dex in the mean metallicity of an old population away from the metallicity of the training set.  For comparison, current modelling of age and metallicity variation along the red sequence of SDSS galaxies suggests a $\sim0.3$~dex change in metallicity across 4 magnitudes of galaxy luminosity \citep{MB}.   Therefore the biases in photo-z's of faint red-sequence galaxies can be highly significant if the photo-z's are trained exclusively on brighter red-sequence members.
\section{Bias from Age Indeterminacy}
Biases from age estimates errors are computed in the same manner as in the metallicity case. The intrinsic bias from the benchmark 10\,Gyr solar metallicity galaxy, $\delta z_{\sub{10\,Gyr}}$, must be subtracted from the total, $\delta z_{\sub{mod}}$, to arrive at the bias attributable to age mismatching: $\Delta z_{\sub{age}} = z_{\sub{mod}} -  z_{\sub{p}} - \delta z_{\sub{10\,Gyr}} = \delta z_{\sub{mod}} - \delta z_{\sub{10\,Gyr}}$.
\begin{figure}[ht!]
	\epsscale{.406}\plotone{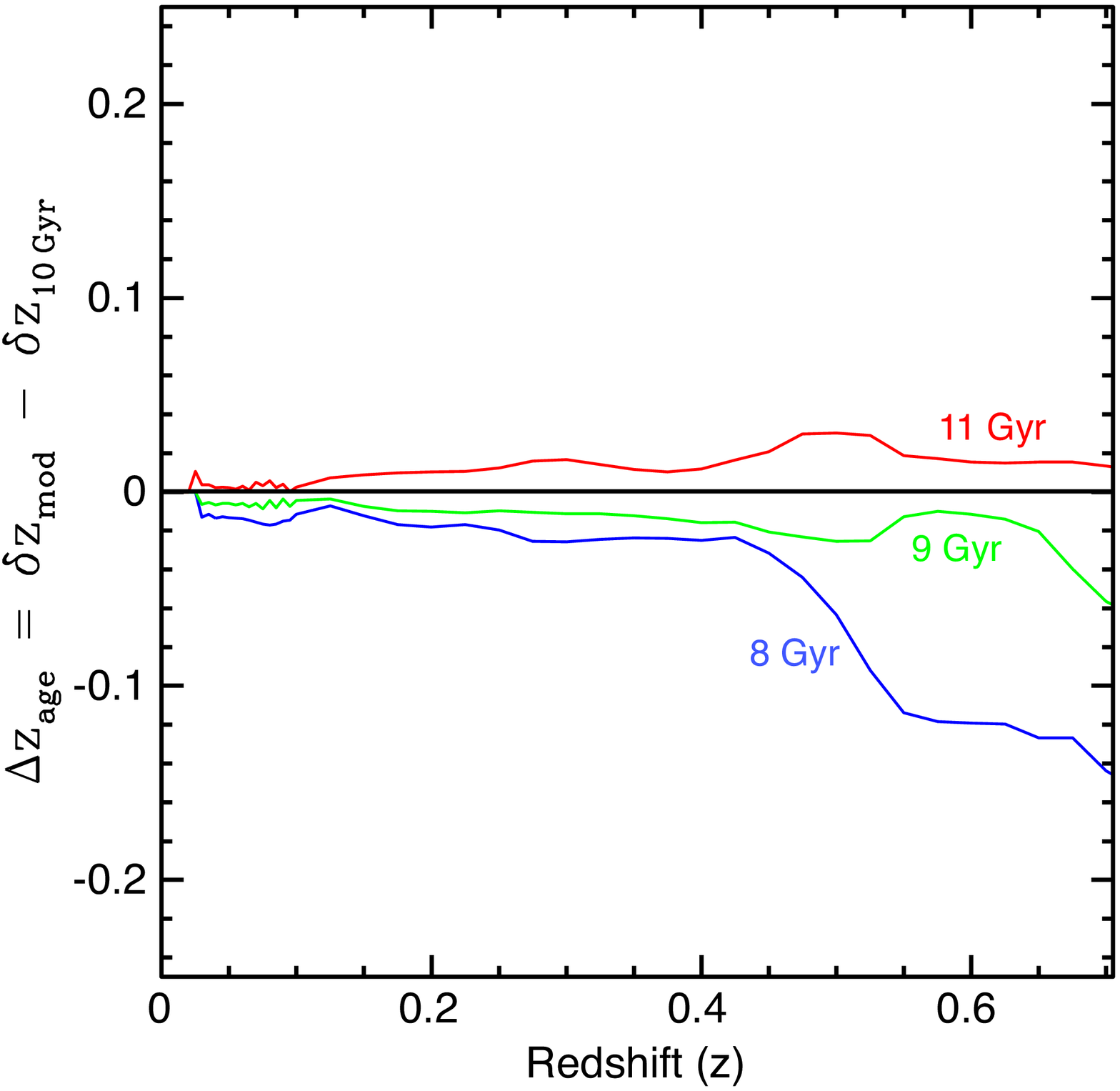}
	\epsscale{.406}\plotone{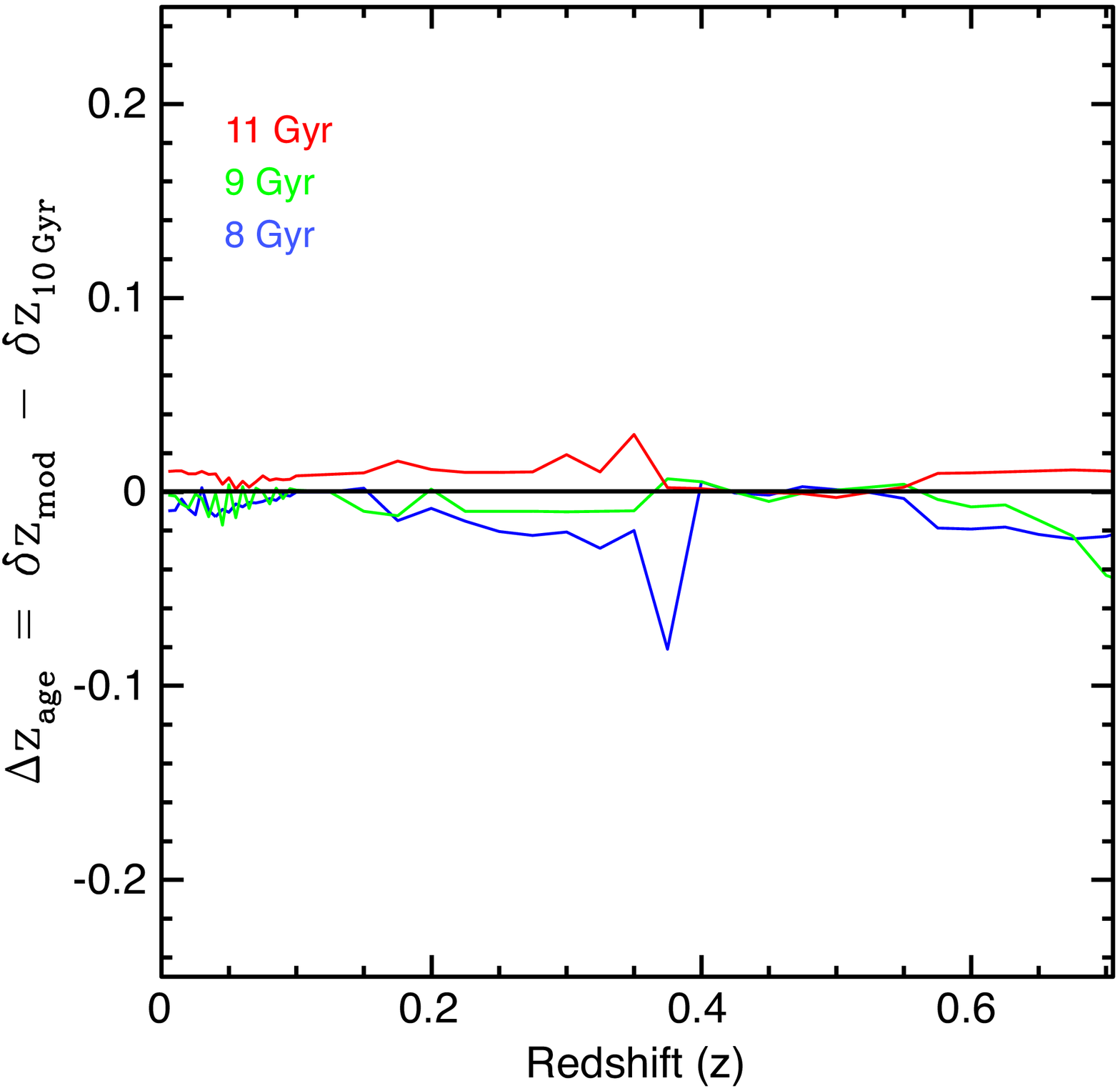}
	\caption{\small Shown above are the age-induced biases for solar metallicity models with $0 \leq 	 z \leq .7$ for the HDFN (left) and SDSS (right) filter sets. Model ages are distinguished by 			color. As with the metallicity-induced bias, a reanalysis of the HDFN data with NIR data excluded yields plots comparable to the SDSS.}\label{azb}
\end{figure}

Figure~\ref{azb} portrays the induced biases in both filter sets. Compared to the metallicity-induced biases, these show little structure outside of the feature near $z=.4$ for the SDSS filters (also found in the metallicity plots), perhaps suggesting that the photo-z's are more sensitive to metallicity offsets. For the HDFN filter set, the bias only becomes comparable at higher $z$ when the age offset is 2\,Gyr. In both studies, the HDFN sets exhibit a slightly wider range in biases which may mean that extra bands may be more confusing for \lephare if the templates and data are not initially well matched.

\section{Conclusion}
Photometric redshifts resemble the real redshifts only as closely as the training sets mimic the objects of interest. In the cases of AGN activity and metallicity variation, we find that even very small mismatches between the mean photometric target and the training set can induce photo-z biases large enough to corrupt significantly the validity of derived cosmological parameters.  In particular, we find that a metallicity shift of $\sim0.003$~dex in an old population, or contamination of any galaxy spectrum with $\sim 0.2\%$ AGN flux, is sufficient to induce a $10^{-3}$ bias in photo-z. Miscalculating the age of a galaxy by 2\,Gyr may also result in similar a bias. In a real survey, we can expect differences far larger than this between the faint photometry survey targets and the bright spectroscopic targets used for photo-z training.

While simple models are used here to study redshift biases, the results indicate a need to develop training sets that encompass the full range of behavior of the photo-z target population.  Our study uses a worst-case situation in that the photo-z algorithm is given no information on the physical effects underlying the bias (AGN or a luminosity-metallicity correlation).  In reality the training sets can include galaxies with AGN of varying brightness and some galaxies of lower luminosity, which would ameliorate the biases we have found.  However the extreme sensitivity of the photo-z to these effects suggests that it will be dangerous to extrapolate from the behavior of bright training-set galaxies to a target photo-z sample that extends to significantly higher redshift and lower luminosity, since even subtle differences in spectral behavior can lead to important biases.  Furthermore we have investigated two known differences between the faint and bright populations, but there may be differences that have not yet been discovered because there are no detailed spectroscopic surveys of the galaxies fainter than 24th magnitude that will comprise the bulk of future weak-lensing surveys.  These results highlight the desirability of training photo-z's with complete spectroscopic surveys to $\sim$25 mag, so that the training set is representative of the galaxies for which photo-z's are being obtained.

\begin{acknowledgments}
This work is supported by grant AST-0607667 from the NSF and DOE grant
DE-FG02-95ER40893.  We are very grateful to Mariangela Bernardi and Gordon Richards for their advice and assistance.  We also thank Stephanie Jouvel for instruction in the use of \Lephare.
\end{acknowledgments}

\end{document}